\documentclass{elsart}
\usepackage{graphics}
\usepackage{epsfig}
\usepackage{amssymb}
\usepackage{cite}
\usepackage{subfigure}

\begin{document}
  \newcommand{\greeksym}[1]{{\usefont{U}{psy}{m}{n}#1}}
  \newcommand{\umu}{\mbox{\greeksym{m}}}
  \newcommand{\udelta}{\mbox{\greeksym{d}}}
  \newcommand{\uDelta}{\mbox{\greeksym{D}}}
  \newcommand{\uOmega}{\mbox{\greeksym{W}}}
  \newcommand{\uPi}{\mbox{\greeksym{P}}}
  \newcommand{\ualpha}{\mbox{\greeksym{a}}}
  \begin{frontmatter}


\vspace*{-11mm}{\it \begin{flushleft} \small
Talk presented at the 10$^{\rm{th}}$ European Symposium on Semiconductor Detectors,\\
June 12-16 2005, Wildbad Kreuth, Germany.
\end{flushleft}
}
\title{A double junction model of irradiated silicon pixel sensors for LHC}

\author[uniz]{V.~Chiochia\corauthref{cor1}}, \ead{vincenzo.chiochia@cern.ch}
\author[jhu]{M.~Swartz},
\author[uniz]{Y.~Allkofer},
\author[purdue]{D.~Bortoletto},
\author[miss]{L.~Cremaldi},
\author[basel]{S.~Cucciarelli},
\author[uniz,psi]{A.~Dorokhov},
\author[uniz,psi]{C.~H\"ormann},
\author[jhu]{D.~Kim},
\author[basel]{M.~Konecki},
\author[psi]{D.~Kotlinski},
\author[uniz,psi]{K.~Prokofiev},
\author[uniz]{C.~Regenfus},
\author[psi]{T.~Rohe},
\author[miss]{D.~A.~Sanders},
\author[purdue]{S.~Son},
\author[uniz]{T.~Speer}

\corauth[cor1]{Corresponding author}
\address[uniz]{Physik Institut der Universit\"at Z\"urich-Irchel, 8057 Z\"urich, Switzerland}
\address[jhu]{Johns Hopkins University, Baltimore, MD 21218, USA}
\address[purdue]{Purdue University, West Lafayette, IN 47907, USA}
\address[miss]{University of Mississippi, University, MS 38677, USA}
\address[basel]{Institut f\"ur Physik der Universit\"at Basel, 4056 Basel, Switzerland}
\address[psi]{Paul Scherrer Institut, 5232 Villigen PSI, Switzerland}
\begin{abstract}
In this paper we discuss the measurement of charge collection in irradiated
silicon pixel sensors and the comparison with a detailed simulation.
The simulation implements a model of radiation damage by including two
defect levels with opposite charge states and trapping of charge carriers.
The modeling proves that a doubly peaked electric field generated by the
two defect levels is necessary to describe the data and excludes a description
based on acceptor defects uniformly distributed across the sensor bulk.
In addition, the dependence
of trap concentrations upon fluence is established by comparing the measured
and simulated profiles at several fluences and bias voltages.
\end{abstract}

\end{frontmatter}

\section{Introduction}
The CMS experiment, currently under construction at the Large Hadron Collider
(LHC) will include a silicon pixel detector~\cite{CMS_Tracker_TDR} to allow tracking in the region closest
to the interaction point. The detector will be a key component for reconstructing interaction
vertices and heavy quark decays in a particulary harsh environment, characterized by 
a high track multiplicity and heavy irradiation. 
The innermost layer, located at only 4 cm from the beam line, is expected
to be exposed to an equivalent fluence of $3 \times 10^{14}$~n$_{\rm eq}$/cm$^2$/yr
at full luminosity.

In these conditions, the response of the silicon sensors during the detector
operation is of great concern. It is well understood that the intra-diode electric fields in these detectors vary linearly in depth reaching a maximum value at the p-n junction.  The linear behavior is a consequence of a constant space charge density, $N_{\rm eff}$, caused by thermodynamically ionized impurities in the bulk material.  It is well known that the detector characteristics are affected by radiation exposure, but it is generally assumed that the same picture is valid after irradiation.  In fact, it is common to characterize the effects of irradiation in terms of a varying effective charge density.
In~\cite{Chiochia:2004qh} we have proved that this picture does not provide a good description of
irradiated silicon pixel sensors. In addition, it was shown that it is possible to adequately describe
the charge collection characteristics of a heavily irradiated silicon detector in terms 
of a tuned double junction model which produces a double peak electric field profile across the sensor.
The modeling is supported by the evidence of doubly peaked electric fields obtained directly from beam
test measurements and presented in~\cite{Dorokhov:2004xk}.
In this paper we apply our model to sensors irradiated to lower fluences 
demonstrating that a doubly peaked electric field is already visible 
at a fluence of $0.5\times$10$^{14}$~n$_{\rm eq}$/cm$^2$. In addition, the dependence
of trap concentrations upon fluence is established by comparing the measured
and simulated profiles at several fluences and bias voltages.

This paper is organized as follows: Section~\ref{sec:technique} describes
the experimental setup, Section~\ref{sec:simulation} describes the carrier transport 
simulation used to interpret the data. The tuning of the double junction model
is discussed in Section~\ref{sec:data_analysis} with the results of the
fit procedure. The conclusions are given in Section~\ref{sec:conclusions}.

\section{Experimental setup\label{sec:technique}}
The measurements were performed in the H2 line of the CERN SPS in 2003/04 using 150-225 GeV pions.  The beam test apparatus is described in~\cite{Dorokhov:2003if}.
A silicon beam telescope \cite{Amsler:2002ta} consisted of four modules each containing two 300~$\umu$m thick single-sided silicon detectors with a strip pitch of 25 $\umu$m and readout pitch of 50~$\umu$m.  The two detectors in each module were oriented to measure horizontal and vertical impact coordinates.  A pixel hybrid detector was mounted between the second and third telescope modules on a cooled rotating stage.  A trigger signal was generated by a silicon
PIN diode. The analog signals from all detectors were digitized in a VME-based readout system by two CAEN (V550) and one custom built flash ADCs. 
The entire assembly was located in an open-geometry 3T Helmholtz magnet that produced a magnetic field parallel or orthogonal to the beam. 
The temperature of the tested sensors was controlled with a Peltier cooler that was capable of operating down to -30$^\circ$C.  The telescope information was used to reconstruct the trajectories of individual beam particles and to achieve a precise determination of the particle hit position
in the pixel detector.  The resulting intrinsic resolution of the beam telescope was about 1~$\umu$m.

The prototype pixel sensors are so-called ``n-in-n'' devices: they are designed to collect charge from n$^+$ structures implanted into n--bulk silicon. All test devices were 22$\times$32 arrays of 125$\times$125~$\mu$m$^2$ pixels having a sensitive area of 2.75$\times$4~mm$^2$.  The substrate was 285~$\mu$m thick, n-doped, diffusively-oxygenated silicon of orientation $\langle111\rangle$, resistivity of about 3.7~k$\Omega\cdot$cm and oxygen concentration in the order of $10^{17}$ cm$^{-3}$.  Individual sensors were diced from fully processed wafers after the deposition of under-bump metalization and indium bumps.  A number of sensors were irradiated at the CERN PS with 24 GeV protons. The irradiation was performed without cooling or bias. The delivered proton fluences scaled to 1 MeV neutrons by the hardness factor 0.62~\cite{Lindstrom:2001ww} were $0.5\times$10$^{14}$~n$_{\rm eq}$/cm$^2$, $2\times$10$^{14}$~n$_{\rm eq}$/cm$^2$ and $5.9\times$10$^{14}$~n$_{\rm eq}$/cm$^2$. All samples were annealed for three days at 30$^\circ$C. In order to avoid reverse annealing, the sensors were stored at -20$^\circ$C after irradiation and kept at room temperature only for transport and bump bonding. All sensors were bump bonded to PSI30/AC30 readout chips \cite{Meer:PSI30} which allow analog readout of all 704 pixel cells without zero suppression.  The PSI30 settings were adjusted to provide a linear response to input signals ranging from zero to more than 30,000 electrons.

\section{Sensor simulation\label{sec:simulation}}

A detailed sensor simulation was implemented, including a physical modeling of irradiation effects in silicon.
Our simulation, {\sc pixelav}~\cite{Swartz:2003ch,Swartz:CMSNote,Chiochia:2004qh}, incorporates the following elements: an accurate model of charge deposition by primary hadronic tracks (in particular to model delta rays); a realistic 3-D electric field map resulting from the simultaneous solution of Poisson's Equation, continuity equations, and various charge transport models; an established model of charge drift physics including mobilities, Hall Effect, and 3-D diffusion; a simulation of charge trapping and the signal induced from trapped charge; and a simulation of electronic noise, response, and threshold effects.  A final step reformats the simulated data into test beam format so that it can be processed by the test beam analysis software.

The effect of irradiation was implemented in the simulation by including
two defect levels in the forbidden silicon bandgap with opposite
charge states and trapping of charge carriers. The model, similar to one proposed in~\cite{Eremin:2002wq},
is based on the Shockley-Read-Hall statistics 
and produces an effective space charge density $\rho_\mathrm{eff}$ from the trapping 
of free carriers in the leakage current.  
The effective charge density is related to the occupancies and densities of traps as follows,
\begin{equation}
\rho_\mathrm{eff} = e\left[N_Df_D-N_Af_A\right] + \rho_\mathrm{dopants}   \label{eq:rhodef}
\end{equation}
where: $N_D$ and $N_A$ are the densities of donor and acceptor trapping states, respectively; $f_D$ and $f_A$ are the occupied fractions of the donor and acceptor states, respectively, and $\rho_\mathrm{dopants}$ is the charge density due to ionized dopants.
Each defect level is characterized by an electron and hole trapping cross section,
$\sigma^D_{e/h}$ and $\sigma^A_{e/h}$, for the donor and acceptor trap, respectively, and by an activation energy, $E_D$ and
$E_A$ for the donor and acceptor trap, respectively. 
 \begin{figure}[hbt]
%
%
  \begin{center}
    \resizebox{0.8\linewidth}{!}{\includegraphics{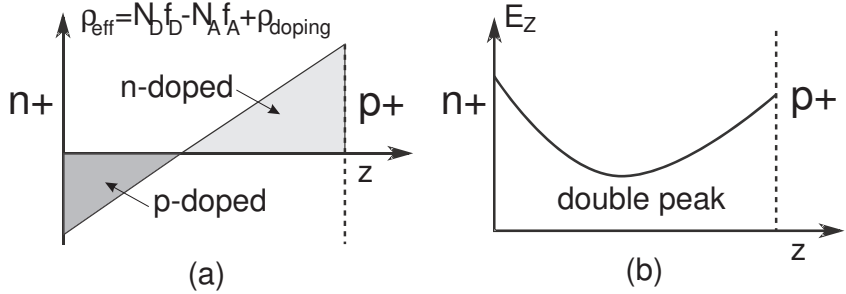}}
  \caption{An illustrative sketch of the double trap model for a reverse biased device.}
  \label{fig:evl_sketch} 
  \end{center}
\end{figure}
An illustrative sketch of the model is shown in Fig.~\ref{fig:evl_sketch}. Trapping of the mobile carriers from the generation-recombination current produces a net positive space charge density near the p$^+$ backplane and a net negative space charge density near the n$^+$ implant as shown in Fig.~\ref{fig:evl_sketch}(a).  Since positive space charge density corresponds to n-type doping and negative space charge corresponds to p-type doping, there are p-n junctions at both sides of the detector.  The electric field in the sensor follows from a simultaneous solution of Poisson's equation and the continuity equations.  The resulting $z$-component of the electric field is shown in Fig.~\ref{fig:evl_sketch}(b).  It varies with an approximately quadratic dependence upon $z$ having a minimum at the zero of the space charge density and maxima at both implants.
A more detailed description of the double junction model and its implementation
can be found in~\cite{Chiochia:2004qh}.

\section{Data analysis\label{sec:data_analysis}}
\begin{figure}[hbt]
%
%
  \begin{center}
    \resizebox{0.8\linewidth}{!}{\includegraphics{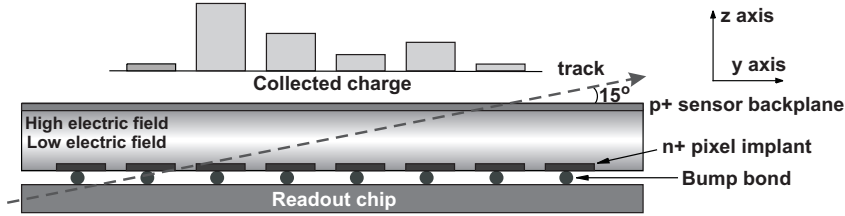}}
  \caption{The grazing angle technique for determining charge collection profiles.  The charge measured by each pixel along the $y$ direction samples a different depth $z$ in the sensor.}
  \label{fig:fifteen_deg} 
  \end{center}
\end{figure}

%
Charge collection across the sensor bulk was measured using the ``grazing angle technique''~\cite{Henrich:CMSNote}.  As is shown in Fig.~\ref{fig:fifteen_deg}, the surface of the test sensor is oriented by a small angle (15$^\circ$) with respect to the pion beam.  A large sample of data is collected with zero magnetic field and at a temperature of $-10^\circ$C.  The charge measured by each pixel along the $y$ direction samples a different depth $z$ in the sensor.  Precise entry point information from the beam telescope is used to produce finely binned charge collection profiles.

\begin{figure}[thb]
%
%
  \begin{center}
    \resizebox{0.75\linewidth}{!}{\includegraphics{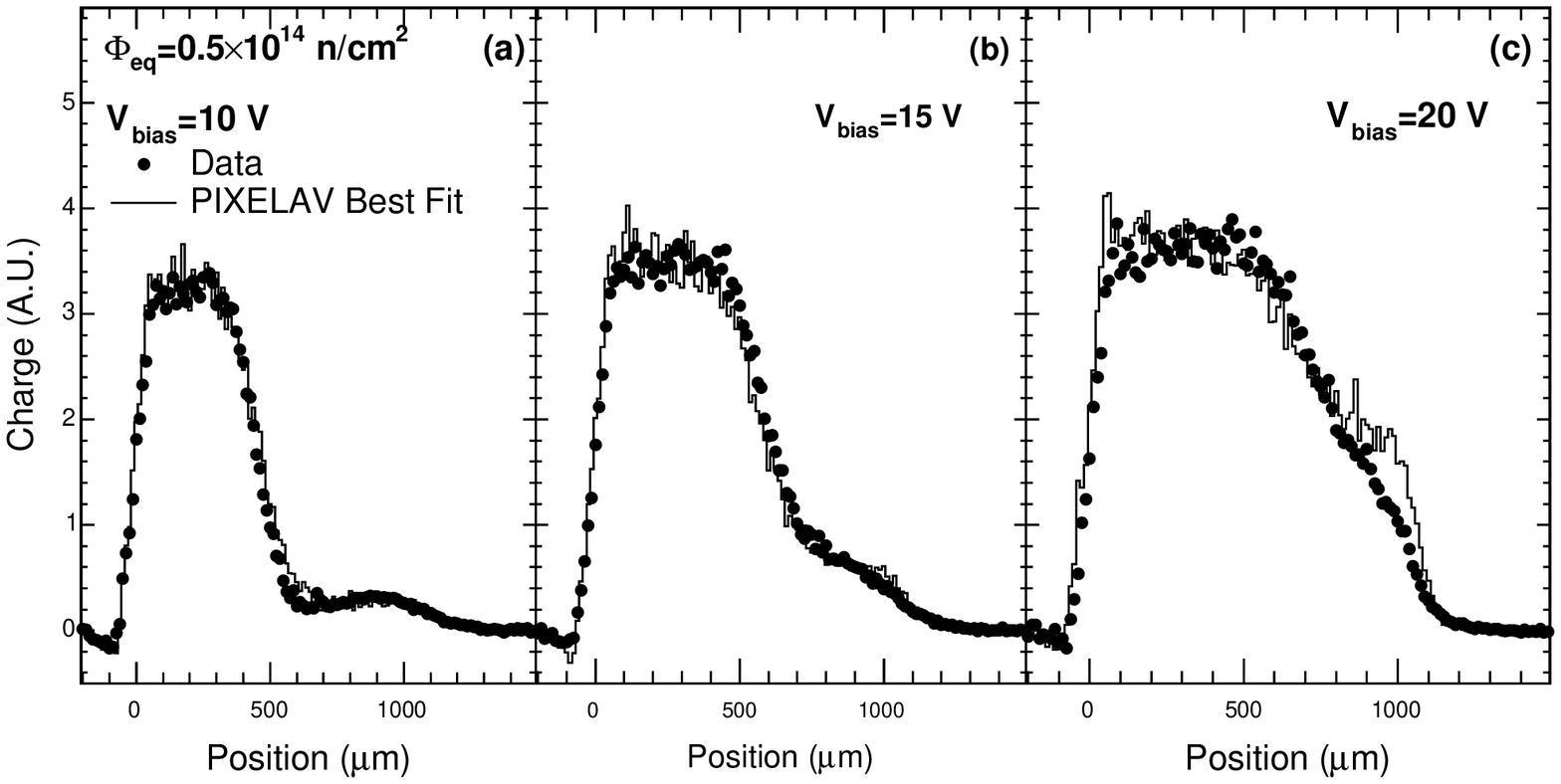}}
    \resizebox{\linewidth}{!}{\includegraphics{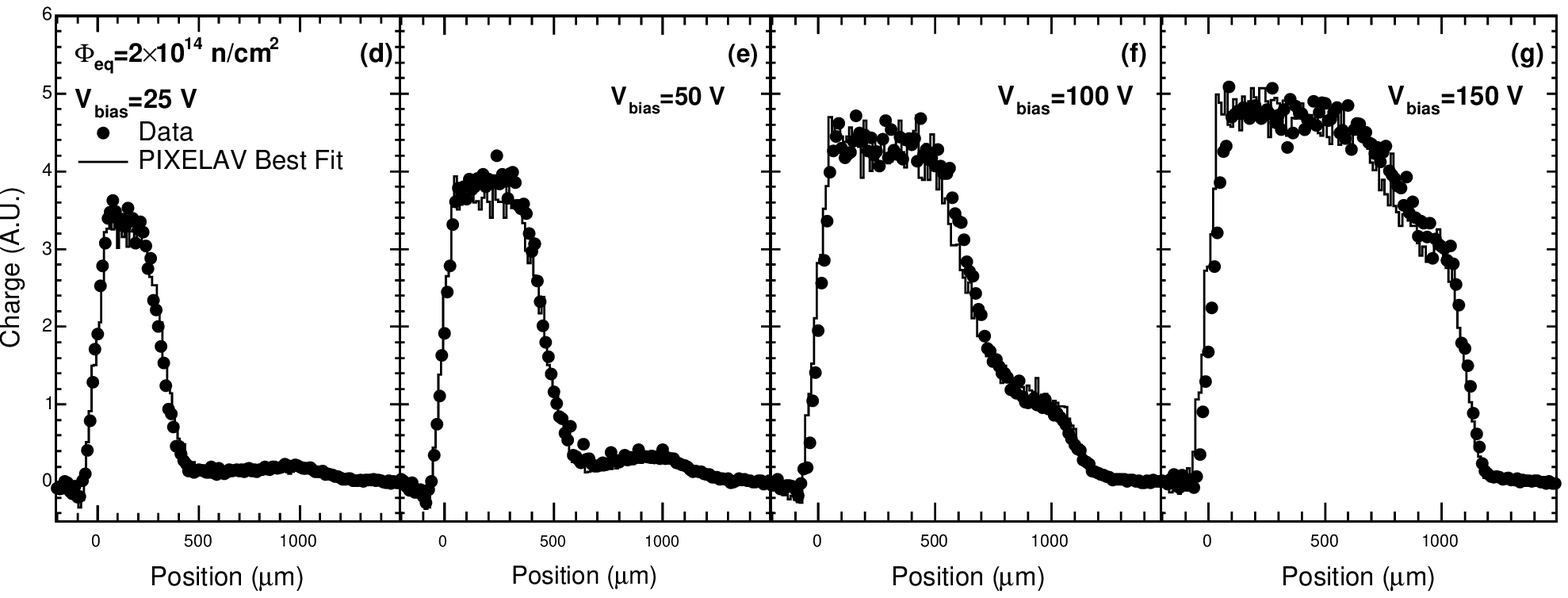}}
  \caption{Measured (full dots) and simulated (histogram) charge collection profiles for a sensor irradiated 
to a fluence of $\Phi = 0.5\times10^{14}$~n$_{\rm eq}$/cm$^2$ (a-c) and of $\Phi = 2\times10^{14}$~n$_{\rm eq}$/cm$^2$ (d-g),
and operated at several bias voltages.}
  \label{fig:summary_2N} 
  \end{center}
\end{figure}

The charge collection profiles for a sensor irradiated to a fluence of
$\Phi = 0.5\times10^{14}$~n$_{\rm eq}$/cm$^2$ and $\Phi = 2\times10^{14}$~n$_{\rm eq}$/cm$^2$ 
and operated at several bias voltages are presented in Fig.~\ref{fig:summary_2N}(a-c)
and Fig.~\ref{fig:summary_2N}(d-g), respectively. The measured profiles,
shown as solid dots, are compared to the simulated profiles, shown as histograms.
The two trap model has six free parameters 
($N_D$, $N_A$, $\sigma_e^D$, $\sigma_h^D$, $\sigma_e^A$, $\sigma_h^A$) 
that can be adjusted.  The activation energies are kept fixed to the values of~\cite{Eremin:2002wq}.  
Additionally, the electron and hole 
trapping rates, $\Gamma_e$ and $\Gamma_h$, are uncertain at the 30\% level due 
to the fluence uncertainty and possible annealing of the sensors.  
They are treated as constrained parameters. The donor concentration of the starting
material is set to $1.2\times10^{12}$~cm$^{-3}$ corresponding to a full depletion
voltage of about 70~V for an unirradiated device.
The parameters of the double junction model were systematically varied and the agreement 
between measured and simulated charge collection profiles was judged subjectively.
The procedure was repeated at the each fluence and the optimal
parameter set was chosen when agreement between measured and simulated profiles
was achieved for all bias voltages.

\begin{figure}[thb]
%
%
  \begin{center}
    \mbox{
      \subfigure[]{\scalebox{0.30}{
	  \epsfig{file=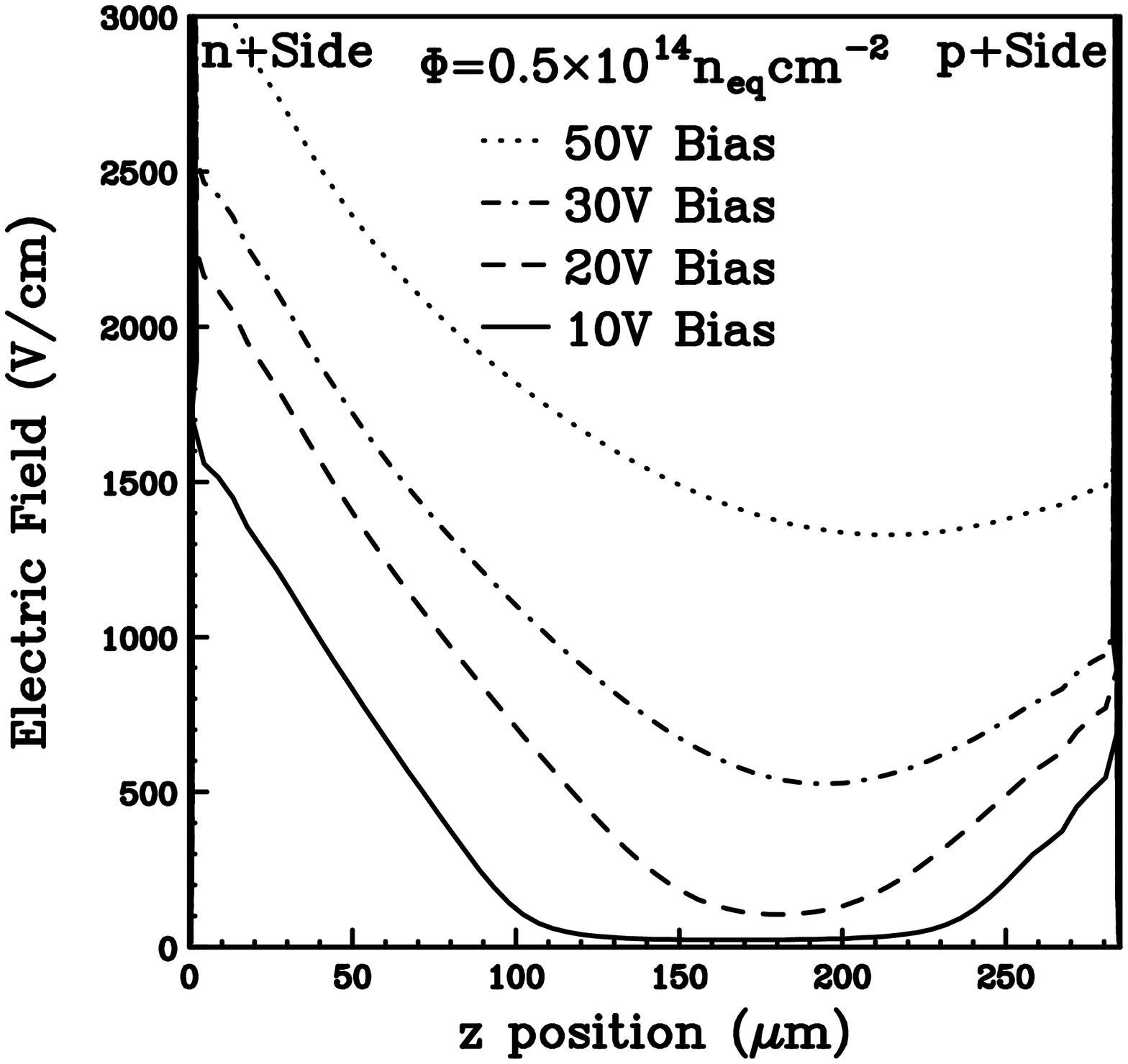,width=\linewidth,angle=90}
	  \label{fig:E_z_profile_05N}
      }}
      \subfigure[]{\scalebox{0.30}{
	  \epsfig{file=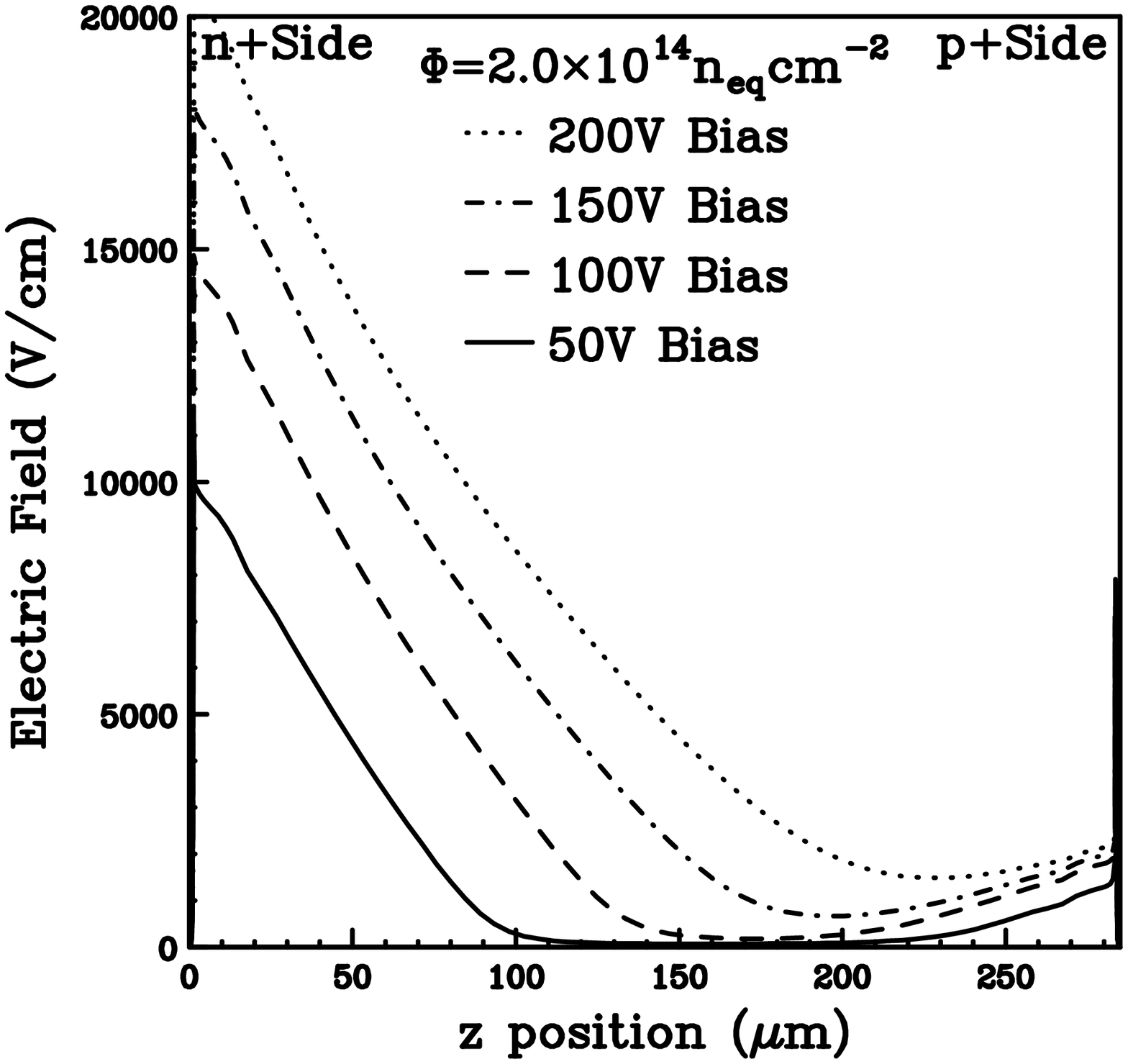,width=\linewidth,angle=90}
	  \label{fig:E_z_profile_2N}
      }}
      \subfigure[]{\scalebox{0.30}{
	  \epsfig{file=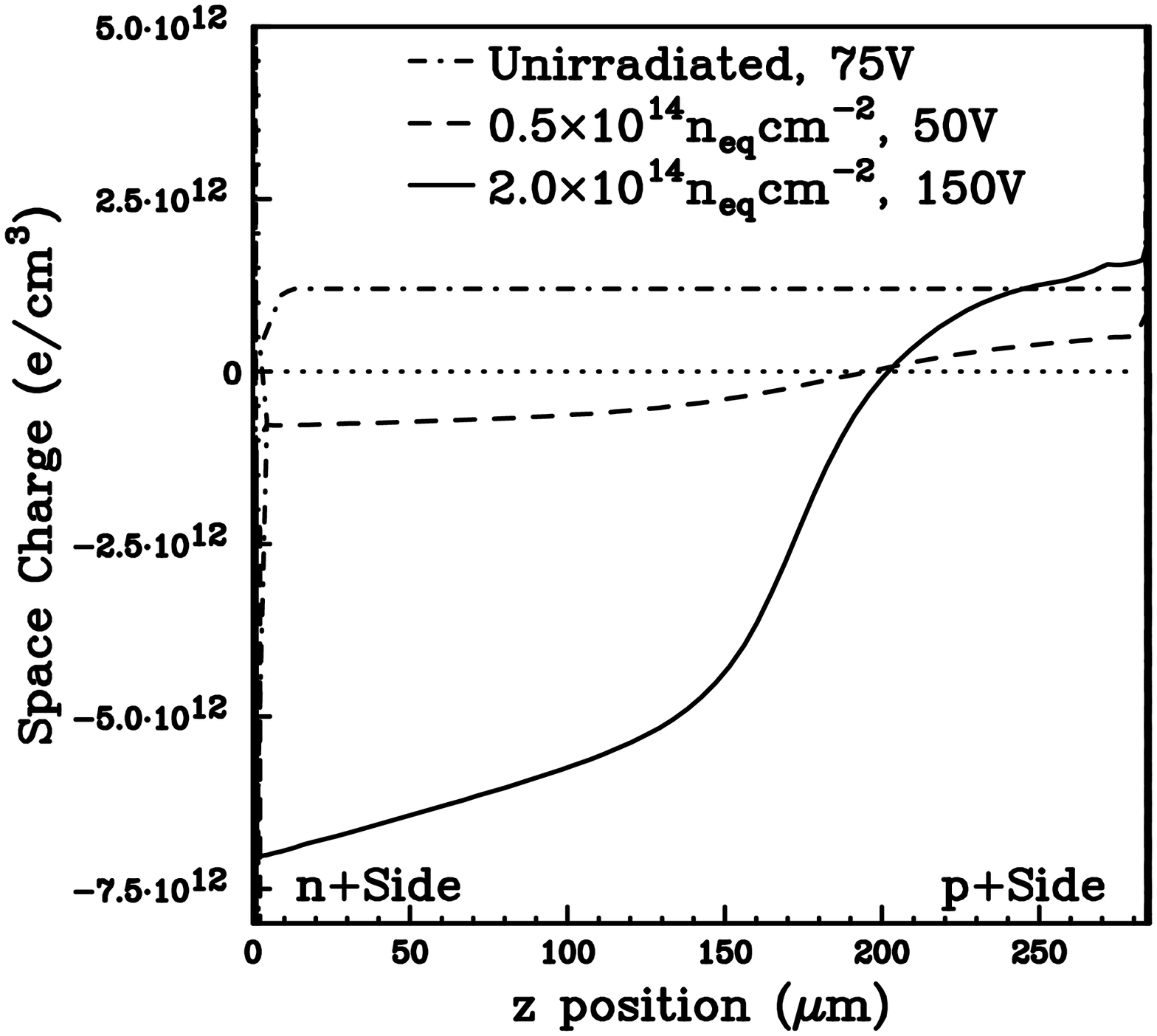,width=\linewidth,angle=90}
	  \label{fig:spacecharge}
      }}
    }
    \caption{The $z$-component of the simulated electric field resulting from the model best fit is shown as a function of $z$
      for a sensor irradiated to a fluence of $\Phi = 0.5\times10^{14}$~n$_{\rm eq}$/cm$^2$ (a) and $\Phi = 2\times10^{14}$~n$_{\rm eq}$/cm$^2$ (b). (c) Space charge density as a function of the $z$ coordinate for different fluences and bias voltages.}
  \end{center}
\end{figure}

The simulation describes the measured charge collection profiles well both
in shape and normalization. 
In particular,the ``wiggle'' observed at low bias voltages is also nicely described.  
The relative signal minimum near $y=700\ \mu$m (see Fig.~\ref{fig:summary_2N}) corresponds to the minimum of the electric field $z$-component, $E_z$, where both electrons and holes travel only short distances before trapping.  This small separation induces only a small signal on the n$^+$ side of the detector.  At larger values of $y$, $E_z$ increases causing the electrons drift back into the minimum where they are likely to be trapped.  However, the holes drift into the higher field region near the p$^+$ implant and are more likely to be collected.  The net induced signal on the n$^+$ side of the detector therefore increases and creates the local maximum seen near $y=900\ \mu$m.
The $z$-component of the simulated electric field, $E_z$, is plotted as a function of $z$ in Fig.~\ref{fig:E_z_profile_05N} and  
Fig.~\ref{fig:E_z_profile_2N} for $\Phi = 0.5\times10^{14}$~n$_{\rm eq}$/cm$^2$ and $\Phi = 2\times10^{14}$~n$_{\rm eq}$/cm$^2$,
respectively. The field profiles have minima near the midplane of the detector
and maxima at the detector implants as discussed in Section~\ref{sec:simulation}. Figure~\ref{fig:E_z_profile_05N} shows that a 
double peak electric field is necessary to describe the measured charge collection profiles 
even at the lowest measured fluence, usually referred to as close to the ``type inversion point''.
The dependence of the space charge density upon the $z$ coordinate is shown in Fig.~\ref{fig:spacecharge}.
Before irradiation the sensor is characterized by a constant and positive space charge density of $1.2\times10^{12}$~cm$^{-3}$
across the sensor bulk.
After a fluence of $0.5\times10^{14}$~n$_{\rm eq}$/cm$^2$ the device shows a negative
space charge density of about $-1\times10^{12}$~cm$^{-3}$ for about 70\% of its thickness, a compensated
region corresponding to the $E_z$ minimum and a positive space charge density close to the backplane. 
The increase of the space charge density  upon $z$ is not linear due to the varying
charge carrier mobilities across the bulk and to the requirement of a constant current density. 

The model parameters obtained with the best fit procedure are shown in 
Table~\ref{tab:model_values}\footnote{The comparison of the measured and simulated profiles 
at $\Phi = 6\times10^{14}$~n$_{\rm eq}$/cm$^2$ can be found in~\cite{Chiochia:2004qh}.}. 
We observe that the donor trap concentration increases more rapidly with fluence
than does the acceptor trap concentration. The ratio between acceptor
and donor trap concentrations is 0.76 at the lowest fluence and decreases 
to 0.40 at $6\times$10$^{14}$~n$_{\rm eq}$/cm$^2$. In addition, the fits exclude 
a linear dependence of the trap concentrations with the irradiation fluence.
At $\Phi = 6\times$10$^{14}$~n$_{\rm eq}$/cm$^2$ the cross section ratio 
$\sigma_h/\sigma_e$ is set to 0.25 for both donor and acceptor traps
while at lower fluences we find $\sigma^{A}_h/\sigma^{A}_e = 0.25$ and 
$\sigma^{D}_h/\sigma^{D}_e = 1$ for the acceptor and donor traps, respectively.
%
%
\begin{table}[h]
\begin{center}
\begin{tabular}{cccccc}
\hline
$\Phi$                          & $N_A$     & $N_D$     & $\sigma^{A/D}_e$ & $\sigma^{A}_h$ & $\sigma^{D}_h$ \\
(n$_{\rm eq}$/cm$^2$) & (cm$^{-3}$) & (cm$^{-3}$) & (cm$^2$)         & (cm$^2$)       & (cm$^2$)       \\
\hline
$0.5\times10^{14}$ & $0.19\times10^{15}$ & $0.25\times10^{15}$ & $6.60\times10^{-15}$ & $1.65\times10^{-15}$ & $6.60\times10^{-15}$ \\
$2\times10^{14}$   & $0.68\times10^{15}$ & $1.0\times10^{15}$ & $6.60\times10^{-15}$ & $1.65\times10^{-15}$ & $6.60\times10^{-15}$ \\
$5.9\times10^{14}$ & $1.60\times10^{15}$ & $4.0\times10^{15}$ & $6.60\times10^{-15}$ & $1.65\times10^{-15}$ & $1.65\times10^{-15}$ \\
\hline
\end{tabular}
\caption{Double trap model parameters extracted from the fit to the data.}
\end{center}
\label{tab:model_values}
\end{table}

\section{Conclusions\label{sec:conclusions}}
%
In this paper we show that a model of irradiated silicon sensors based on two
defect levels with opposite charge states and trapping of charge carriers can
be tuned using charge collection measurements and 
provides a good description of the measured charge collection profiles
in the fluence range from $0.5\times$10$^{14}$~n$_{\rm eq}$/cm$^2$
to $6\times$10$^{14}$~n$_{\rm eq}$/cm$^2$. 

The model produces an electric
field profile across the sensor that has maxima at the implants and
a minimum near the detector midplane. This corresponds to  
negative space charge density near the n$^+$ implant and
and positive space charge density near the p$^+$ backplane. 
We find that it is necessary to decrease the ratio of acceptor concentration
to donor concentration as the fluence increases. This causes the electric
field profile to become more symmetric as the fluence increases.

Given the extracted electric field and space charge density profiles
we suggest that the correctness and the physical significance 
of effective doping densities determined from capacitance-voltage measurements
are quite unclear. In addition, we remark
that the notion of partly depleted silicon sensors after irradiation 
is inconsistent with the measured charge collection profiles and 
with the observed doubly peaked electric fields.

The charge-sharing behavior and resolution functions of many detectors are sensitive to the details of the internal electric field.  A known response function is a key element of any reconstruction procedure.  A working effective model will permit the detailed response of these detectors to be tracked as they are irradiated in the next generation of accelerators.

\section*{Acknowledgments}
We gratefully acknowledge Silvan Streuli from ETH Zurich and Fredy Glaus from PSI for
their immense effort with the bump bonding, Federico Ravotti, Maurice Glaser and Michael Moll from CERN for
carrying out the irradiation, Kurt B\"osiger from the Z\"urich University workshop for the mechanical
construction, Gy\"orgy Bencze and Pascal Petiot from CERN for the H2 beam line support
and, finally, the whole CERN-SPS team.


\bibliographystyle{elsart-num}    


\bibliography{refs}             

\end{document}